\documentclass[aps,prl,reprint,showpacs,longbibliography]{revtex4-2} 
\usepackage{graphicx} 
\usepackage{amsmath}  
\usepackage{amssymb}  
\usepackage[colorlinks=true,linkcolor=blue,urlcolor=blue,citecolor=blue]{hyperref} 
\usepackage{xr}
\usepackage{tikz}
\usepackage{float}

\usetikzlibrary{decorations.markings}
\begin{document}

\title{Electronic conductivity in anharmonic crystals: \\ phonon dephasing in the electron-phonon interaction}
\author{Mingran Kong}
\author{Bartomeu Monserrat}
\affiliation{Department of Materials Science and Metallurgy, University of Cambridge,
27 Charles Babbage Road, Cambridge CB3 0FS, United Kingdom}
\date{\today} 

\begin{abstract}
The electron-phonon interaction underpins many material properties, for example the conductivity of metals and the optoelectronic response of semiconductors. First principles calculations of the electron-phonon interaction
are a powerful tool to quantitatively describe many of these properties in increasingly complex materials. However, one key assumption of all calculations is that phonons have infinite lifetimes, an approximation that may break down when anharmonic phonon-phonon interactions are strong. In this work, we present a theory for the interaction of electrons with finite-lifetime phonons experiencing dephasing. Using a first principles implementation of the theory, we find that anharmonic dephasing dramatically enhances electron-phonon scattering rates in metallic MgB$_2$. Microscopically, phonon-phonon interactions create new scattering channels that increase the phase space available for electron-phonon scattering. As a result, anharmonic dephasing strongly suppresses conductivity in MgB$_2$, bringing the calculated values substantially closer to experiment within the Boltzmann transport equation framework.
This example establishes the importance of finite phonon lifetimes in the evaluation of electron-phonon scattering, and the microscopic mechanism suggests that anharmonic dephasing could play an important role in the conductivity of many metals. More broadly, our theory and first principles implementation of anharmonic dephasing in the electron-phonon interaction provides a solid foundation to explore this regime in other materials.
\end{abstract}


\maketitle


The electron-phonon interaction underpins a vast array of physical properties of materials, from transport\,\cite{Payne1983, Payne1983_2, Cui2015, Waldecker2016, Maldonado2020} to superconductivity\,\cite{Maxwell1950, Reynolds1950, Bardeen1957, Eliashberg1960, McMillan1968}. The first principles description of electron-phonon coupling is a key ingredient in the quantitative understanding of material properties\,\cite{Profeta2012, Monserrat2014, Antonius2014, Errea2015, Faber2015, Antonius2016, Monserrat2016, Coulter2018, Chen2020, Miglio2020, Yang2021, Lee2023, Abramovitch2024, Li2024, Dolui2024, Garmroudi2025}, and plays an increasingly important role in the discovery of new materials, for example in the area of hydride superconductors\,\cite{Duan2014}.  

The current first principles framework for the evaluation of electron-phonon coupling makes the key assumption that phonons have infinite lifetimes. 
However, anharmonic phonon-phonon scattering, which is present in all materials and dominates in materials containing light elements\,\cite{Ravichandran2020} or at high temperatures\,\cite{Skelton2016}, leads to finite phonon lifetimes, or equivalently, to phonon dephasing. 
The assumption of infinite phonon lifetimes can be justified if phonon dephasing occurs on timescales much longer than those relevant to electron-phonon scattering, in which case electron-phonon and phonon-phonon processes are temporally separated.
However, this assumption may break down in materials with strong phonon anharmonicity, and the role of anharmonic phonon dephasing on the electron-phonon interaction remains an open question.

\begin{figure}[t]
\centering
\includegraphics[width=0.7\linewidth]{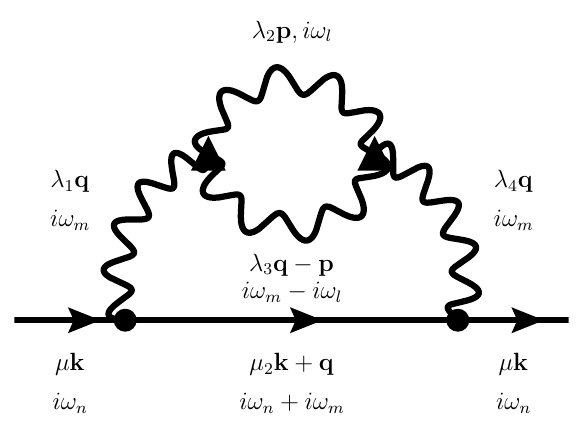}
\caption{Feynman digram for electron-anharmonic-phonon scattering}
\label{fig:feynman-diagram}
\end{figure}

In this work, we present a formula for the electron-phonon scattering rate in the presence of anharmonic dephasing. Our approach combines the Fan-Migdal electron-phonon interaction with three-phonon anharmonic interactions, as captured by the Feynman diagram in Fig.\,\ref{fig:feynman-diagram}. This theory represents the lowest-order correction to electron-phonon coupling arising from finite phonon lifetimes, and qualitatively captures electrons interacting with phonons that are undergoing phonon-phonon interactions. We also present a first principles implementation of anharmonic dephasing in the electron-phonon interaction, which provides a powerful platform to study this phenomenon in a wide range of materials. Additionally, in a companion work we provide a detailed step-by-step derivation of the formula for anharmonic dephasing of the electron-phonon interaction and we describe it in the wider context of electron-phonon interactions in anharmonic crystals\,\cite{companion}.

Finally, we also evaluate the magnitude of anharmonic dephasing on the electron-phonon coupling of metallic magnesium diboride (MgB$_2$), a material that is known to exhibit strong anharmonicity\,\cite{Liu2001}. 
We find that finite phonon lifetimes dramatically modify electron-phonon scattering rates, and directly lead to a significant reduction of the electronic conductivity of MgB$_2$, bringing calculated values in closer agreement with experiment. This result highlights the potentially large impact of finite phonon lifetimes in material properties that are derived from the electron-phonon interaction.

We consider a Hamiltonian including first-order electron-phonon interactions and anharmonic three-phonon interactions: 
\begin{equation}
\begin{split}
\hat{H} &= \sum_{\mu \mathbf{k}} \epsilon_{\mu \mathbf{k}} \hat{c}_{\mu \mathbf{k}}^\dagger \hat{c}_{\mu \mathbf{k}} 
+ \sum_{\lambda \mathbf{q}} \omega_{\lambda \mathbf{q}} 
\left(\hat{b}_{\lambda \mathbf{q}}^\dagger \hat{b}_{\lambda \mathbf{q}} + \frac{1}{2}\right)\\
&
+ \frac{1}{\sqrt{N}} \sum_{\mu_2\mu_1 \mathbf{k}} \sum_{\lambda \mathbf{q}} g_{\mu_2 \mu_1 \lambda}^{\mathbf{k} \mathbf{q}}
\hat{c}^\dagger_{\mu_2\mathbf{k}+\mathbf{q}} \hat{c}_{\mu_1 \mathbf{k}} \hat{A}_{\lambda \mathbf{q}} \\
&
+ \frac{1}{3!\sqrt{N}}
\sum_{\substack{\lambda_1\lambda_2\lambda_3\\\mathbf{q}_1\mathbf{q}_2\mathbf{q}_3}}
\phi_{\lambda_1\lambda_2\lambda_3}^{\mathbf{q}_1\mathbf{q}_2\mathbf{q}_3}
\hat{A}_{\lambda_1\mathbf{q}_1}
\hat{A}_{\lambda_2\mathbf{q}_2}
\hat{A}_{\lambda_3\mathbf{q}_3},
\end{split}
\label{eq:hamiltonian}
\end{equation}
where $\hat{c}_{\mu \mathbf{k}}$ ($\hat{c}_{\mu \mathbf{k}}^\dagger$) are the electron annihilation (creation) operators for an electron with momentum $\mathbf{k}$, band $\mu$, and energy $\epsilon_{\mu\mathbf{k}}$; $\hat{A}_{\lambda \mathbf{q}} = \hat{b}_{\lambda \mathbf{q}} + \hat{b}_{\lambda \mathbf{q}}^\dagger$ are the phonon field operators for mode $\lambda$ with momentum $\mathbf{q}$ and frequency $\omega_{\lambda\mathbf{q}}$, comprising of the associated phonon creation $\hat{b}_{\lambda \mathbf{q}}^\dagger$ and annihilation $\hat{b}_{\lambda \mathbf{q}}$ operators;
$g_{\mu_2\mu_1\lambda}^{\mathbf{k}\mathbf{q}}$ is the electron-phonon coupling strength; $\phi_{\lambda_1\lambda_2\lambda_3}^{\mathbf{q}_1\mathbf{q}_2\mathbf{q}_3}$ is the three-phonon coupling strength; and
$N$ is the number of unit cells in the system.
The Hamiltonian in Eq.\,(\ref{eq:hamiltonian}) describes an interacting system with phonons coupling with electrons and with other phonons, with the latter three-phonon term representing the anharmonic phonon contribution that we include in this work to go beyond the standard description of electron-phonon systems. 

Within many-body perturbation theory, the lowest-order term exhibiting anharmonic phonon dephasing is captured by the Feynman diagram depicted in Fig.\,\ref{fig:feynman-diagram} (see companion work). This diagram leads to an electron-anharmonic-phonon self-energy that is a combination of the Fan-Midgal electron-phonon self-energy and the phonon-phonon bubble self-energy. 
The associated scattering rate under the self-energy relaxation time approximation is given by (see companion work):
\begin{equation}
\Gamma_{\mu\mathbf{k}}^{\mathrm{el}\text{-}\mathrm{anh}\text{-}\mathrm{ph}} = 2\Gamma_{\mu\mathbf{k}}^\mathrm{(1e1a)} + \Gamma_{\mu\mathbf{k}}^\mathrm{(2e)} + \Gamma_{\mu\mathbf{k}}^\mathrm{(2a)}.
\label{eq:gamma-main}
\end{equation}
In this expression, the one-phonon emitted and one-phonon absorbed (1e1a) term can be written as:

\begin{widetext}
\begin{equation}
\begin{split}
\Gamma_{\mu\mathbf{k}}^\mathrm{(1e1a)} = 
\frac{\pi}{N^2} 
\sum_{\lambda_2\lambda_3}
\sum_{\mathbf{q}\mathbf{p}}
&\left| 
    \sum_{\lambda_1} 
    \Bigg(
        \frac{
            g_{\mu_1\mu\lambda_1}^{\mathbf{k}\mathbf{q}}
            \phi_{\lambda_1\lambda_2\lambda_3}^{\mathbf{q}\mathbf{p}(\mathbf{q}-\mathbf{p})*}
        }{
            \omega_{\lambda_2} - \omega_{\lambda_3} - \omega_{\lambda_1} + i\eta
        }
        -
        \frac{
            g_{\mu_1\mu\lambda_1}^{\mathbf{k}\mathbf{q}}
            \phi_{\lambda_1\lambda_2\lambda_3}^{\mathbf{q}\mathbf{p}(\mathbf{q}-\mathbf{p})*}
        }{
            \omega_{\lambda_2} - \omega_{\lambda_3} + \omega_{\lambda_1} + i\eta
        }
    \Bigg)
\right|^2 \times \\
&\delta({\omega_n} - \omega_{\lambda_2} + \omega_{\lambda_3} - \epsilon_{\mu_1})
\Big[
\mathcal{N}_{\lambda_2}\mathcal{N}_{\lambda_3} + \mathcal{N}_{\lambda_2}
\Big],
\end{split}
\label{eq:gamma-1e1a}
\end{equation}
the two-phonon emitted (2e) term can be written as:
\begin{equation}
\begin{split}
\Gamma_{\mu\mathbf{k}}^\mathrm{(2e)} = 
&\frac{\pi}{N^2} 
\sum_{\lambda_2\lambda_3}
\sum_{\mathbf{q}\mathbf{p}}
\left| 
    \sum_{\lambda_1} 
    \Bigg(
        \frac{
            g_{\mu_1\mu\lambda_1}^{\mathbf{k}\mathbf{q}}
            \phi_{\lambda_1\lambda_2\lambda_3}^{\mathbf{q}\mathbf{p}(\mathbf{q}-\mathbf{p})*}
        }{
            \omega_{\lambda_2} + \omega_{\lambda_3} - \omega_{\lambda_1} + i\eta
        }
        -
        \frac{
            g_{\mu_1\mu\lambda_1}^{\mathbf{k}\mathbf{q}}
            \phi_{\lambda_1\lambda_2\lambda_3}^{\mathbf{q}\mathbf{p}(\mathbf{q}-\mathbf{p})*}
        }{
            \omega_{\lambda_2} + \omega_{\lambda_3} + \omega_{\lambda_1} + i\eta
        }
    \Bigg)
\right|^2 \times \\
&\delta({\omega_n} - \omega_{\lambda_2} - \omega_{\lambda_3} - \epsilon_{\mu_1})
\Big[
(\mathcal{N}_{\lambda_2}+\mathcal{N}_{\lambda_3}+1)(f_{\mu_1}-1) + \mathcal{N}_{\lambda_2}\mathcal{N}_{\lambda_3}
\Big],
\end{split}
\label{eq:gamma-2e}
\end{equation}
and the two-phonon absorbed (2a) term can be written as:
\begin{equation}
\begin{split}
\Gamma_{\mu\mathbf{k}}^{\mathrm{(2a)}} =
-\frac{\pi}{N^2}
\sum_{\lambda_2\lambda_3}
\sum_{\mathbf{q}\mathbf{p}}
&\left|
\sum_{\lambda_1}
\Bigg(
\frac{
g_{\mu_1\mu\lambda_1}^{\mathbf{k}\mathbf{q}}
\phi_{\lambda_1\lambda_2\lambda_3}^{\mathbf{q}\mathbf{p}(\mathbf{q}-\mathbf{p})}
}{
\omega_{\lambda_2} + \omega_{\lambda_3} - \omega_{\lambda_1} + i\eta
}
-
\frac{
g_{\mu_1\mu\lambda_1}^{\mathbf{k}\mathbf{q}}
\phi_{\lambda_1\lambda_2\lambda_3}^{\mathbf{q}\mathbf{p}(\mathbf{q}-\mathbf{p})}
}{
\omega_{\lambda_2} + \omega_{\lambda_3} + \omega_{\lambda_1} + i\eta
}
\Bigg)
\right|^2 \times \\
\delta(\omega_n + \omega_{\lambda_2} + &\omega_{\lambda_3} - \epsilon_{\mu_1})
\Big[
(\mathcal{N}_{\lambda_2}+\mathcal{N}_{\lambda_3}+1)(f_{\mu_1}-1) +
\mathcal{N}_{\lambda_2}\mathcal{N}_{\lambda_3} + \mathcal{N}_{\lambda_2} + \mathcal{N}_{\lambda_3} + 1
\Big],
\end{split}
\label{eq:gamma-2a}
\end{equation}
\end{widetext}
where 
$f_\mu$ and $\mathcal{N}_\lambda$ denote the electron and phonon occupation numbers, respectively.




Qualitatively,
these formulae represent electrons coupling with phonons dressed by a phonon-phonon bubble diagram. Therefore, with the introduction of anharmonicity the bare phonon states are corrected and dephased, and electrons scatter with these dressed phonons. 
Through this mechanism, the phonon-phonon interaction governs the likelihood of phonons entering the incoherent regime and also dictates which states are more likely to indirectly couple with electrons. Consequently, it may significantly alter the microscopic scattering mechanisms, especially in materials where phonons experience strong dephasing.
This should be contrasted with electron-two-phonon scattering from higher-order electron-phonon perturbation theory\,\cite{Lee2020}, where the electron undergoes sequential scattering by two phonons through intermediate electronic states and the phonons retain infinite lifetimes. 

\begin{figure*}[t]
\centering
\includegraphics[width=\textwidth]{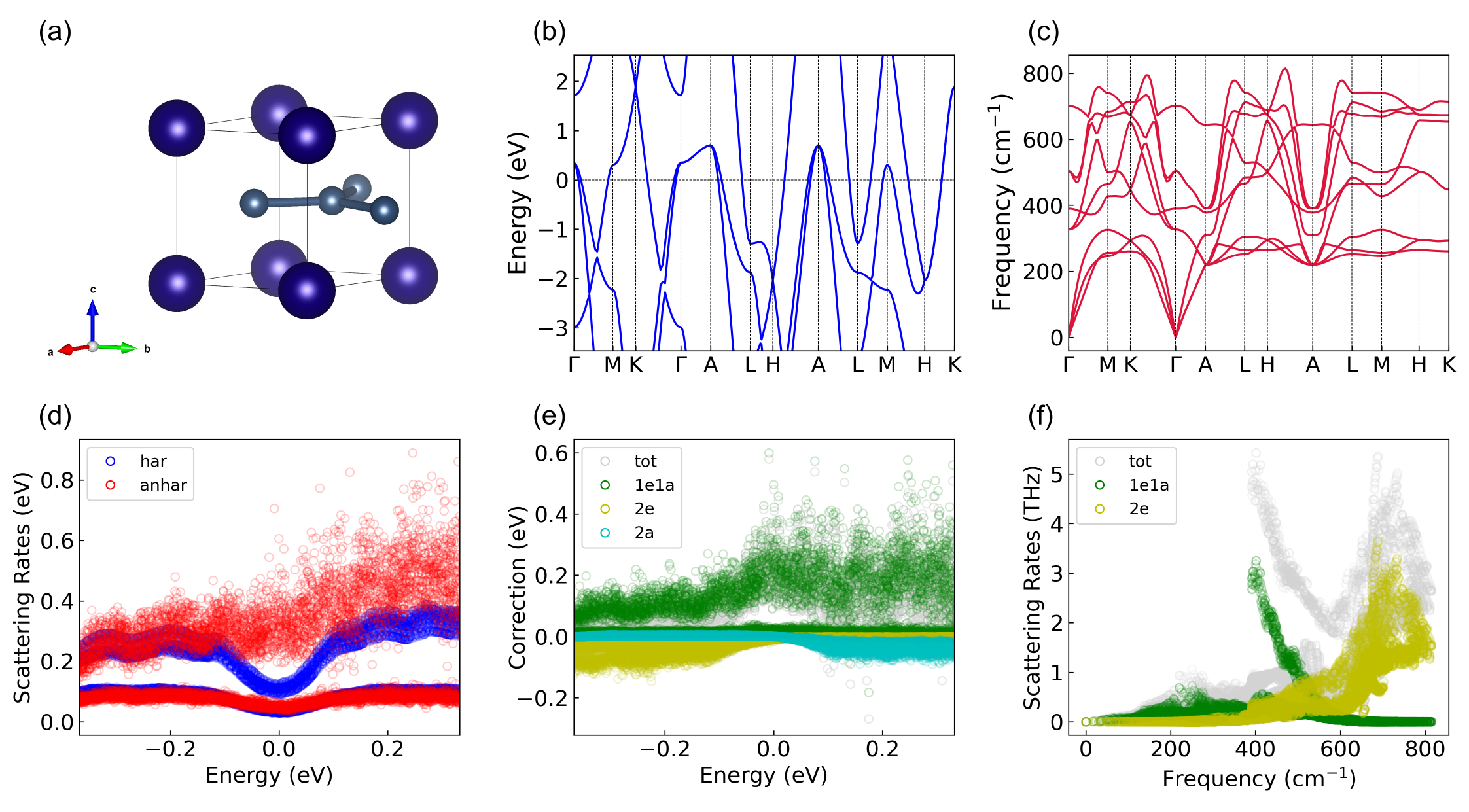}
\caption{
MgB$_2$ results, including (a) crystal structure;
(b) electronic band structure;
(c) phonon dispersion;
(d) electron-phonon scattering rates in blue and electron-phonon plus electron-anharmonic-phonon scattering rates in red at $300$\,K;
(e) process-resolved electron-anharmonic-phonon correction versus electron energy; (f) process-resolved three-phonon scattering rates versus phonon frequencies.
}
\label{fig:scattering-rates}
\end{figure*}


We have implemented the formulae in Eqs.\,(\ref{eq:gamma-main})-(\ref{eq:gamma-2a}) within a first principles framework in our in-house code \texttt{DaoQuantum}, full details of which will be reported elsewhere. In our implementation (see companion work), we obtain the key electron, phonon, electron-phonon, and phonon-phonon parameters from \texttt{Quantum ESPRESSO}, use \texttt{EPW} to generate the electron-phonon matrix elements in the Wannier basis, and \texttt{ShengBTE} to generate the third-order interatomic force constants. We then combine all of these ingredients within \texttt{DaoQuantum} to evaluate the role of anharmonic dephasing in the electron-phonon scattering rates, and to calculate the electronic conductivity incorporating these anharmonic scattering rates.


We apply our methodology to study the electronic conductivity of MgB$_2$ in its normal metal phase, and the corresponding crystal structure is presented in Fig.\,\ref{fig:scattering-rates}(a). 
Figure~\ref{fig:scattering-rates}(b) shows the electronic structure of MgB$_2$, which exhibits multiple bands crossing the Fermi level, leading to multiple electron-phonon scattering channels which, for example, drive the presence of multiple superconducting gaps in MgB$_2$ at low temperature\,\cite{Kong2001,Liu2001}. 
As shown in Fig.\,\ref{fig:scattering-rates}(c), the phonon dispersion of MgB$_2$ continuously spans a broad range of frequencies, facilitating energy- and momentum-conserving phonon-phonon scatterings that underpin the anharmonic behaviour of this material. These features of the electron and phonon dispersions suggest 
that electron-anharmonic-phonon scattering could be significant in MgB$_2$.


To facilitate the comparison of the electron-phonon coupling properties calculated using the standard approach, harmonic phonons and linear electron-phonon coupling, with electron-phonon coupling incorporating anharmonic phonon dephasing, we will refer to the former as ``harmonic electron-phonon coupling'' and to the latter as ``anharmonic electron-phonon coupling''. 

The harmonic electron-phonon scattering rates for MgB$_2$ as a function of the electron energy around the Fermi level are presented in Fig.\,\ref{fig:scattering-rates}(d) in blue.
These scattering rates split into two distinct ``bands'', one below about $0.1$\,eV, and the other spanning a wider energy range between $0.1$-$0.4$\,eV and exhibiting a pronounced drop near the Fermi level. The two distinct scattering rates emerge from two different microscopic mechanisms: the low-energy band arises from weak coupling between phonons and out-of-plane $\pi$ bands, while the high-energy band arises from strong coupling of the two-dimensional $\sigma$ bands to the in-plane E$_{2g}$ phonon mode\,\cite{Kong2001,Liu2001}. The drop of the harmonic scattering rate near the Fermi level likely arises from restrictions on available phonon states participating in the scattering.

The anharmonic electron-phonon scattering rates for MgB$_2$ are also shown in Fig.\,\ref{fig:scattering-rates}(d) in red, and they also exhibit two distinct ``bands''. The lower energy feature closely tracks the corresponding harmonic one, indicating weak phonon-phonon scattering for the associated modes. Interestingly, the higher energy feature is significantly modified when anharmonic phonon-phonon interactions are included. First, the anharmonic scattering rates span a broader energy range compared to the harmonic counterparts, driven by the strong anharmonicity in MgB$_2$, where the conventional single-phonon contribution is modified by various two-phonon combinations. 
Second, and in stark contrast to the harmonic case, there is no drop in the anharmonic scattering rate near the Fermi level, indicating an increase in the phase space available to electron-phonon scattering mediated by the phonon-phonon scatterings. This is an interesting feature, suggesting that anharmonic corrections could play a dominant role in the transport properties, and even superconductivity, of many materials. 


Figure~\ref{fig:scattering-rates}(e) shows the different electron-anharmonic-phonon scattering mechanisms as a function of the electron energy around the Fermi level, and Fig.\,\ref{fig:scattering-rates}(f) shows the energy-conserving phonon-phonon scattering mechanisms as a function of the harmonic phonon frequency. 
Examining these two figures provides insight into how phonon-phonon scattering intersects with electron-phonon scattering.
The anharmonic electron-phonon correction is primarily driven by the 1e1a-type scattering processes [dark green in Fig.\,\ref{fig:scattering-rates}(e)], which are always positive (see companion work). The associated phonon-phonon mechanism [dark green in Fig.\,\ref{fig:scattering-rates}(f)] exhibits a peak around $400$\,cm$^{-1}$ which smoothly decreases towards $600$\,cm$^{-1}$, forming a distinct ``waterfall'' pattern,
a feature that suggests the presence of resonant scatterings between multiple phonon valleys at an energy around $400$\,cm$^{-1}$\,\cite{Delaire2011}.
Indeed, the phonon dispersion in Fig.\,\ref{fig:scattering-rates}(c) shows several valleys, highlighting those around the $\Gamma$ point, that facilitate momentum- and energy-conserving scattering processes.
An additional smaller phonon-phonon scattering peak appears between $200$\,cm$^{-1}$ to $300$\,cm$^{-1}$ in Fig.\,\ref{fig:scattering-rates}(f) (dark green), likely a consequence of the extremely flat phonon bands in this energy range [Fig.\,\ref{fig:scattering-rates}(c)].

The 2e phonon-phonon scattering rate also exhibits a pronounced peak centred around $700$\,cm$^{-1}$ [olive green in Fig.\,\ref{fig:scattering-rates}(f)], arising from the high phonon density of states in this spectral region.
However, the associated two-phonon–emission contribution to electron–anharmonic-phonon scattering is markedly weaker than the 1e1a channel, as shown in Fig.\,\ref{fig:scattering-rates}(e) (olive green circles).
This observation suggests that the pronounced 1e1a contribution stems predominantly from a simultaneous strong anharmonicity and electron-phonon coupling of the E$_{2g}$ phonon mode, whose frequency ranges from $400$\,cm$^{-1}$ at the $A$ point up to around $600$\,cm$^{-1}$ at the K point.
Considering the central role that the E$_{2g}$ phonon mode plays in mediating Cooper pairing in superconducting MgB$_2$, our results suggest the possibility that anharmonic dephasing in the electron-phonon interaction may significantly renormalize the pairing interaction in the superconducting regime\,\cite{Nagamatsu2001}, a question we leave for future work.


\begin{figure}[t]
\centering
\includegraphics[width=0.38\textwidth]{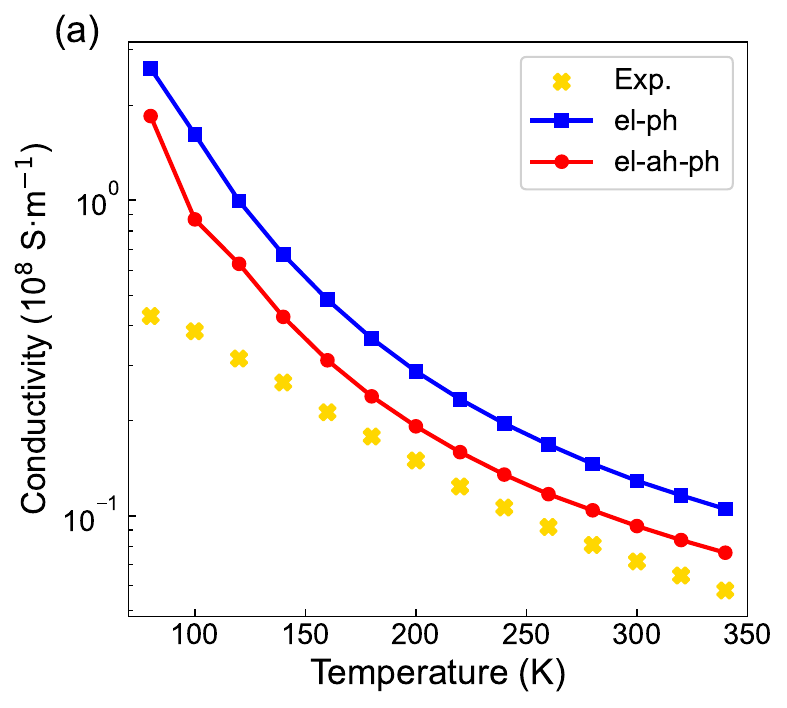}
\caption{
Electrical conductivity of MgB$_2$ from $80$\,K to $340$\,K calculated using harmonic electron-phonon coupling (blue) and anharmonic electron-phonon coupling (red), and compared to expreimental results from Ref.\,\cite{Sologubenko2002} (yellow crosses).
}
\label{fig:transport}
\end{figure}


Using the electron-phonon scattering rates incorporating anharmonic dephasing, we calculate the electrical conductivity of MgB$_2$ from $80$\,K and $340$\,K from first principles (see Supplemental Material\,\cite{SM}), and compare it to the standard harmonic calculation and experiment, as summarized in Fig.\,\ref{fig:transport}.
Amoug these results, the electrical conductivities at the harmonic level are $1.62 \times 10^{8}$\,Sm$^{-1}$ and $1.33 \times 10^{7}$\,Sm$^{-1}$ at $100$\,K and $300$\,K, respectively.
Interestingly, the electrical conductivities including anharmonic dephasing dramatically decrease to $8.70 \times 10^{7}$\,Sm$^{-1}$ and $9.47 \times 10^{6}$\,Sm$^{-1}$ at $100$\,K and $300$\,K, respectively. This decrease is driven by the increase in scattering rates when anharmonic dephasing is included. The anharmonic-corrected conductivities are significantly closer to the experimental values of $3.57 \times 10^{7}$\,Sm$^{-1}$ and $7.14 \times 10^{6}$\,Sm$^{-1}$ at $100$\,K and $300$\,K, respectively. The lower value of the experimental conductivities could be attributed to extrinsic effects, like electron-defect scattering, that are not included in our calculations.
These results illustrate that finite phonon lifetimes are essential to describe electron transport in MgB$_2$.
Other effects, such as thermal expansion, phonon-frequency renormalization, and non-adiabatic corrections\,\cite{Wang2026OppositeImpact}, may further affect the absolute conductivity, but do not alter the conclusion that anharmonic dephasing is an important contribution that should be investigated in a wider range of compounds.


In summary, we present a theory and first principles implementation of the impact of anharmonic phonon-phonon dephasing into the description of electron–phonon interactions. 
We apply this theory to metallic MgB$_2$, 
finding that anharmonic dephasing substantially modifies electron–phonon coupling near the Fermi level by opening new electron-phonon scattering channels mediated by phonon-phonon interactions. The enhanced anharmonic electron-phonon coupling leads to a significant suppression in the calculated electrical conductivity compared to the harmonic counterpart, resulting in better agreement with experiment. 
This example highlights the critical role of phonon dephasing in accurately modeling transport properties in anharmonic systems. The formalism is general and the first principles implementation robust, so they form a solid platform to study a wider range of properties related to electron-phonon coupling in materials exhibiting lattice anharmonicity. 

Looking ahead, MgB$_2$ is a well-known phonon-mediated superconductor\,\cite{Nagamatsu2001}, so it would be interesting to explore the role that anharmonic dephasing could have on its superconducting properties. Another interesting future direction would be the extension of the theory, for example to strongly anharmonic systems where phonon-phonon interactions cannot be treated perturbatively, which could be accomplished using variational techniques to define effective phonons renormalized by the strong anharmonicity.
In this context, emerging data-driven compression\,\cite{Luo2024} and tensor-decomposition approaches\,\cite{Luo2025} for electron-phonon and anharmonic phonon interactions could help reduce the cost of dense interpolation and make applications to more complex anharmonic materials feasible.


\begin{acknowledgements}
M.K. and B.M. acknowledge financial support from the Gianna Angelopoulos Programme for Science, Technology, and Innovation. B.M. also acknowledges support from a UKRI Future Leaders Fellowship [MR/V023926/1].
The computational resources were provided by the UK National Supercomputing Service ARCHER2 and by the UK Materials and Molecular Modelling Hub, which is partially funded by EPSRC [EP/P020194], and access for both was obtained via the UKCP consortium and funded by EPSRC [EP/X035891/1].
\end{acknowledgements}

\bibliography{ref} 

\end{document}


\title{Supplemental Material: \\
Electronic conductivity in anharmonic crystals: phonon dephasing in the electron-phonon interaction}
\author{Mingran Kong}
\author{Bartomeu Monserrat}
\affiliation{Department of Materials Science and Metallurgy, University of Cambridge,
27 Charles Babbage Road, Cambridge CB3 0FS, United Kingdom}
\maketitle

\section{Iterative solution of the Boltzmann transport equation}

Starting from the linearized Boltzmann transport equation in the presence of a weak
electric field $E$,
\begin{equation}
-e\,v^{\beta}_{n\mathbf{k}}\,
\frac{\partial f^{0}_{n\mathbf{k}}}{\partial \varepsilon_{n\mathbf{k}}}
=
\sum_{m}
\int\!\frac{d^3q}{\Omega_{\mathrm{BZ}}}
\Big[
\tau^{-1}_{m\mathbf{k+q}\rightarrow n\mathbf{k}}\,
\partial_{E_\beta} f_{m\mathbf{k+q}}
-
\tau^{-1}_{n\mathbf{k}\rightarrow m\mathbf{k+q}}\,
\partial_{E_\beta} f_{n\mathbf{k}}
\Big],
\label{eq:S_BTE_start}
\end{equation}
where $v_{n\mathbf{k}}^\beta$ is the group velocity, $f_{n\mathbf{k}}$ is the occupation number, $\varepsilon_{n\mathbf{k}}$ is the electron energy, and
$\tau^{-1}_{n\mathbf{k}\rightarrow m\mathbf{k+q}}$ denotes the partial
electron-phonon scattering rate including the first-order electron-phonon scattering and anharmonicity-dephasing correction for the transition from the electronic state
$(n,\mathbf{k})$ to $(m,\mathbf{k}+\mathbf{q})$.
These mode-resolved scattering rates can be calculated with the sum of the Fan-Midgal self-energy and our corrections given in the main text from first principles.
We introduce the total scattering rate:
\begin{equation}
\Gamma_{n\mathbf{k}}
=
\sum_{m}
\int\!\frac{d^3q}{\Omega_{\mathrm{BZ}}}\;
\tau^{-1}_{n\mathbf{k}\rightarrow m\mathbf{k+q}} ,
\label{eq:S_Gamma}
\end{equation}
which allows us to rewrite Eq.~\eqref{eq:S_BTE_start} as a linear system for the
field–derivative of the distribution function:
\begin{equation}
\Gamma_{n\mathbf{k}}\,
\partial_{E_\beta} f_{n\mathbf{k}}
-
\sum_{m}
\int\!\frac{d^3q}{\Omega_{\mathrm{BZ}}}\;
\tau^{-1}_{m\mathbf{k+q}\rightarrow n\mathbf{k}}\,
\partial_{E_\beta} f_{m\mathbf{k+q}}
=
-e\,v^{\beta}_{n\mathbf{k}}\,
\frac{\partial f^{0}_{n\mathbf{k}}}{\partial \varepsilon_{n\mathbf{k}}}.
\label{eq:S_BTE_linear_system}
\end{equation}

We go beyond the relaxation-time approximation (RTA) and solve
Eq.\,\eqref{eq:S_BTE_linear_system} iteratively.  Using the RTA expression as an
initial guess:
\begin{equation}
\partial_{E_\beta} f^{(0)}_{n\mathbf{k}}
=
e\,v^{\beta}_{n\mathbf{k}}\,
\tau_{n\mathbf{k}}\,
\frac{\partial f^{0}_{n\mathbf{k}}}{\partial \varepsilon_{n\mathbf{k}}},
\qquad
\tau_{n\mathbf{k}} = \Gamma_{n\mathbf{k}}^{-1},
\label{eq:S_BTE_RTA_guess}
\end{equation}
we update the response function according to
\begin{equation}
\partial_{E_\beta} f^{(i+1)}_{n\mathbf{k}}
=
\frac{1}{\Gamma_{n\mathbf{k}}}
\left[
e\,v^{\beta}_{n\mathbf{k}}\,
\frac{\partial f^{0}_{n\mathbf{k}}}{\partial \varepsilon_{n\mathbf{k}}}
+
\sum_{m}
\int\!\frac{d^3q}{\Omega_{\mathrm{BZ}}}\;
\tau^{-1}_{m\mathbf{k+q}\rightarrow n\mathbf{k}}\,
\partial_{E_\beta} f^{(i)}_{m\mathbf{k+q}}
\right].
\label{eq:S_BTE_iterative}
\end{equation}
The iteration in Eq.\,\eqref{eq:S_BTE_iterative} is repeated until convergence of
$\partial_{E_\beta} f_{n\mathbf{k}}$ is reached.  The converged solution is then
inserted into the conductivity formula:
\begin{equation}
\sigma_{\alpha\beta}
=
-\frac{e}{V_{\mathrm{uc}}}
\sum_{n}
\int\!\frac{d^3k}{\Omega_{\mathrm{BZ}}}\;
v^{\alpha}_{n\mathbf{k}}\,
\partial_{E_\beta} f_{n\mathbf{k}},
\label{eq:S_sigma_fullBTE}
\end{equation}
which yields the full iterative Boltzmann transport equation conductivity.

\section{Convergence}

The harmonic and anharmonic electron-phonon scattering rates are evaluated by numerically sampling the electron and phonon Brillouin zones using uniform grids. Any quantity derived from these scattering rates, such as the electronic conductivity reported in the main text, has to be converged with respect to the Brillouin zone grid sizes. 

\begin{figure}[ht]
    \centering
    \includegraphics[width=\linewidth]{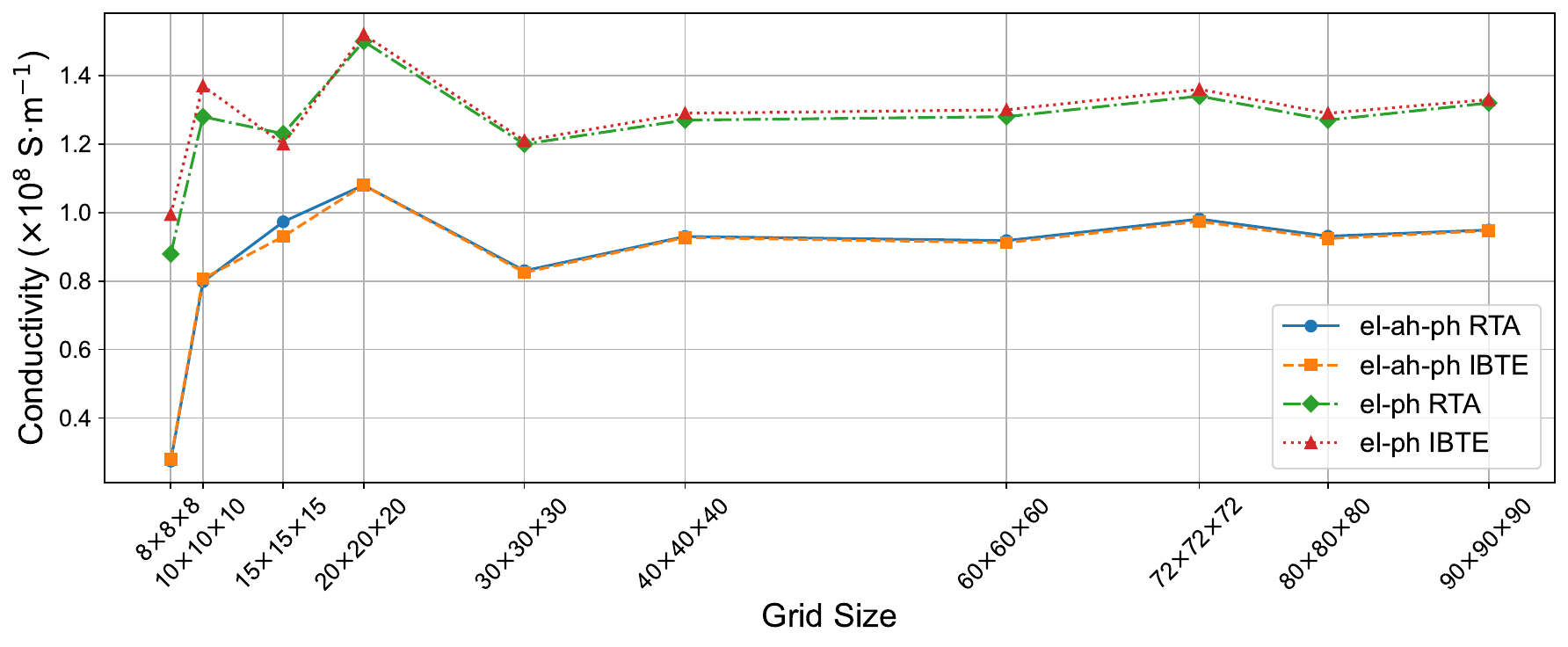}
    \caption{Electronic conductivity of MgB$_2$ as a function of electron and phonon grid sizes in both the harmonic and anharmonic regimes. Solutions using the relaxation time approximation (RTA) and an iterative Boltzmann transport equation (IBTE) solver are reported.}
    \label{fig:convergence_plot}
\end{figure}

We systematically evaluate the harmonic and anharmonic electron-phonon scattering rates of MgB$_2$ using a sequence of electron and phonon Brillouin zone grid sizes from $8\times8\times8$ to $90\times90\times90$, utilizing identical grids for electrons and phonons. We then calculate the resulting harmonic and anharmonic electron–phonon mediated conductivities under the relaxation time approximation (RTA) and an iterative Boltzmann transport equation (IBTE) solver, with the results shown in Fig.\,\ref{fig:convergence_plot}. First, we note that the results obtained within the relaxation time approximation are consistent with those obtained within an interative Boltzmann transport equation solver. In terms of convergence, the computed conductivities rise sharply between $8\times8\times8$ and $20\times20\times20$, reflecting significant under-sampling at coarse grids.
A more modest change appears at $30\times30\times30$, after which all four curves enter a plateau: beyond $40\times40\times40$, variations remain below 7\% for the harmonic electron–phonon mediated conductivity (1.27–1.36\,$\times10^8\,$S\,m$^{-1}$) and below 5\% for the anharmonic electron-phonon mediated conductivity (9.2–9.8\,$\times10^7\,$S\,m$^{-1}$). 
The calculations reported in the main text use a $90\times90\times90$ grid.

\bibliographystyle{apsrev4-2}  
\bibliography{ref}